\documentstyle[aas2pp4,psfig,11pt]{article}

\begin{document}

\title{A variable 0.58--2.44 Hz quasi-periodic oscillation in the
eclipsing and dipping low-mass X-ray binary EXO 0748--676}

\author{Jeroen Homan\altaffilmark{1},
        Peter G. Jonker\altaffilmark{1},
        Rudy Wijnands\altaffilmark{1},
        Michiel van der Klis\altaffilmark{1},
        Jan van Paradijs\altaffilmark{1,2}}

\altaffiltext{1}{Astronomical Institute 'Anton Pannekoek', University
of Amsterdam, and Center for High Energy Astrophysics, Kruislaan 403,
1098 SJ, Amsterdam, The Netherlands}
\altaffiltext{2}{University of Alabama in Huntsville, Department of
Physics, Huntsville, AL 45899}

\begin{abstract}
We report the discovery of a quasi-periodic oscillation (QPO) in data
obtained with the Rossi X-ray Timing Explorer of the dipping and
eclipsing low-mass X-ray binary EXO 0748--676. The QPO had a frequency
between 0.58 and 2.44 Hz changing on time scales of a few days, an rms
amplitude between 8\% and 12\%, and was detected in the persistent
emission, during dips and during type I X-ray bursts. During one
observation, when the count rate was a factor 2 to 3 higher than
otherwise, the QPO was not detected. The strength of the QPO did not
significantly depend on photon energy, and is consistent with being
the same in the persistent emission, both during and outside the dips,
and during type I X-ray bursts. Frequency shifts were observed during
three of the four X-ray bursts. We argue that the QPO is produced by
the same mechanism as the QPO recently found by Jonker et al. (1999)
in 4U 1323--62. Although the exact mechanism is not clear, it is most
likely related to the high inclination of both systems. An orbiting
structure in the accretion disc that modulates the radiation from the
central source seems the most promising mechanism.
\end{abstract}

\keywords{accretion, accretion disks - stars: individual (EXO 0748--676) - stars: neutron - X-rays: stars}

\section{Introduction}
The low-mass X-ray binary (LMXB) EXO 0748--676 was discovered with
{\it EXOSAT} by Parmar et al. (\markcite{pawhgi85}1985). It showed
periodic eclipses, irregular intensity dips, and type I X-ray bursts
(Parmar et al. \markcite{pawhgi86}1986). The eclipses, due to
obscuration of the central X-ray source by the companion star,
occurred at a period of 3.82 hr.  From the eclipse duration an
inclination of 75$^\circ$ to 82$^\circ$ was derived. Dipping activity
is seen at orbital phases $\phi\sim0.8-0.2$ and at $\phi\approx0.65$
($\phi=0$ corresponds to eclipse center). The dips can be as deep as
100\% (Church et al. \markcite{chbado98}1998). It is believed (White
\& Swank \markcite{whsw82}1982; Walter et al. \markcite{waboma82}1982)
that such dips are caused by obscuration of the central X-ray source by a
bulge in the outer accretion disk that is created by the impact of the
gas stream from the companion star, or by the accretion stream itself
if it penetrates the disk further in (Frank, King, \& Lasota
\markcite{frkila87}1987). During the eclipses 4\% of the 2--6 keV
intensity remained, which Parmar et al. (\markcite{pawhgi86}1986)
attributed to the presence of an accretion disk corona (ADC). Using
their `progressive covering' model Church et
al. \markcite{chbado98}(1998) found that $\sim$70\% of the out-of-dip
flux is contributed by this ADC, which they estimated to have a radius
of 0.4--1.5$\times$10$^8$ cm.

Recently, a persistent 1 Hz quasi-periodic oscillation (QPO) was found
with RXTE in the high inclination ($i\approx80^\circ$) dipping LMXB 4U
1323--62 (Jonker, Van der Klis, \& Wijnands \markcite{jovakl99}1999,
hereafter JO99). JO99 concluded that its presence is most likely
related to the high inclination of the system.  In this Letter, we
report the discovery of a variable 0.58--2.44 Hz QPO in EXO 0748--676
with properties similar to those found in 4U 1323--62.
  
\section{Observation and Analysis}
We made use of data obtained with the proportional counter array (PCA;
Jahoda et al. \markcite{jaswgi96}1996) on board the {\it Rossi X-ray
Timing Explorer} ({\it RXTE}; Bradt, Rothschild, \& Swank
\markcite{brrosw93}1993). A log of all observations is given in Table
\ref{table1}. Observations 2 to 7 were originally performed to study
the orbital evolution of the system. Each consisted of five $\sim$2 ks
data intervals that range in orbital phase from $\phi\approx0.9$ to
$\phi\approx0.05$. All observations yielded data in the Standard 1 and
2 modes, which have 1/8 s time resolution in 1 energy channel (2--60
keV) and 16 s time resolution in 129 channels (2--60 keV),
respectively. In addition, data were obtained in modes with at least
2$^{-12}$ s time resolution, in 67 (obs. 1), 32 (obs. 2), 256
(obs. 2--7), or 64 (obs. 8--10) channels, covering the 2--60 keV
range. Data during eclipses were removed from further analysis. X-ray
bursts were studied separately.

Power spectra were created in the energy bands 2--60, 2--5, 5--8,
8--13, and 13--42 keV. The average 1/16--128 Hz power spectra of each
observation were rms normalized, and fitted with a model consisting of
a constant (representing the Poisson level), a power law,
$P\propto\nu^{-\alpha}$ (the noise component), and a Lorentzian (the
QPO). Errors on the parameters were determined using
$\Delta\chi^2=1$. The dependence on photon energy of the QPO strength
was determined by fitting the power spectra in four energy bands,
while keeping the QPO FWHM and frequency, and the power law index
($\alpha$) of the noise component fixed to the values obtained in the
2--60 keV band. Unless stated otherwise the fit parameters are those
in the 2--60 keV band. The 95\% confidence upper limits on the
presence of the QPO and/or noise component were determined by fixing
the FWHM and frequency, and/or by fixing $\alpha$.

For observations 2--6 and 8--10 we compared power spectra obtained in-
and outside the dips. A count rate level was determined for each
single orbit, below which data were assumed to be in a dip. For
observations 8--10 this level was the persistent count rate during
phases were normally no dipping is observed. As mentioned before,
observations 2--6 were taken during phases that normally show dipping,
which is confirmed by increases in hardness. For these observations we
took the post-eclipse count rate as the discriminator, since dipping
usually occurs less after the eclipse.

\section{Results}

Figure \ref{qpo} shows two typical power spectra of EXO 0748--676. A
QPO peak was found in the power spectra of all observations except the
first one. The QPO covered a frequency range of more than a factor 4,
between 0.58$\pm$0.01 Hz (obs. 10) and 2.44$\pm$0.03 Hz (obs. 2), with
rms amplitudes between 8.4\%$\pm$0.4\% and 12.1\%$\pm$0.6\%. Within
each observation ($\sim$1 day) the QPO was only observed in a small
frequency range, never changing more than 15\% in frequency. Table
\ref{table1} gives the fit parameters of all observations in the 2--60
keV energy band.  In observation 1, where the QPO was not detected,
i.e., at least a factor 4 weaker than otherwise, the background
corrected count rate was a factor 2 to 3 higher than in the other
observations.  In the other observations, neither the QPO nor the
noise component show any correlation between their parameters and the
2--60 keV count rate, which varies by $\sim$20\%. However observations
8, 9, and 10, for which (due to the more extended phase coverage) the
persistent count rate could be estimated much better than for the
other observations, show a clear anti-correlation between frequency
and count rate. The QPO peak becomes broader when its frequency
increases (see Figure \ref{ratio}). The ratio $Q$, of frequency over
FWHM of the QPO, is consistent with being constant; a linear fit gives
a $Q$ of 3.5$\pm$0.1, with $\chi^2$/d.o.f.=12.2/8.  The noise
component tends to become steeper as it gets stronger (see Figure
\ref{noise}). The rms of the QPO decreases from $\sim$12\% to
$\sim$8.5\% when the rms of the noise component increases from
$\sim$6\% to $\sim$10.5\%. Although there is considerable scatter
around a linear relation, there seems to be a connection between the
two components (correlation coefficient is --.75). Figure \ref{energy}
shows the photon energy spectra of the QPO in observations 2 to 10 (dip and non-dip data combined). It
is obvious from this figure that they are
quite flat. The energy spectra of the noise component are similar.

All QPO parameters in- and outside the dips are consistent with each
other, except for the QPO amplitude in observation 10, which during
the dips is 12.0\%$\pm$0.8\% rms and outside the dips 7.5\%$\pm$0.4\%
rms. The noise shows only a significant difference in observation 4,
where the dip rms is 9.6$\pm$0.5\% and the non-dip rms is
14.9$\pm$1.4\%.

Four type I X-ray bursts were observed, all outside dips; one in
observations 5 and 8 each, and two in observation 9 (hereafter burst 1
to 4). In 16 s resolution data they had peak count rates (2--60 keV, 3
detectors, background corrected) of 2330$\pm$10, 2240$\pm$10,
2410$\pm$10, and 2110$\pm$10 counts s$^{-1}$, respectively, and they
all lasted $\sim$400 s. In all four bursts the QPO was detected. The
QPO during burst 2 is the only that has width and
frequency that are consistent with those of the QPO outside the
burst. The QPO frequency during burst 1 is 0.1 Hz higher than
that of the QPO outside the burst, and the FWHM is 0.18$\pm$0.05 Hz,
which is considerably smaller than 0.36 Hz. The QPOs during bursts 3
and 4 are both poorly fitted with a Lorentzian. Both peaks appear to
have shifted in frequency during the bursts. The peaked shape in the
power spectrum of burst 4 clearly consists of two small peaks, which,
when fitted with Lorentzians, have frequencies of 0.65$\pm$0.05 Hz and
0.89$\pm$0.02 Hz and a FWHM of $\sim$0.1 Hz.  The QPO in burst 3 is
very asymmetric. The best fit with one Lorentzian at 0.74$\pm$0.02 Hz
and a FWHM of $\sim$0.1 Hz, shows excess power at higher
frequencies. Dynamical power spectra of burst 3 and 4 show that the QPO
profile changed with time.  The QPOs during the four bursts have rms
amplitudes of 9.7\%$\pm$0.9\%, 8.4\%$^{+1.2}_{-1.0}$\%, at least 7\%
(probably 8\%--10\%), and $\sim$10\% (the sum of the rms amplitudes of
the two Lorentzians), respectively, consistent with those found
outside the bursts.

\section{Discussion}

We have discovered a QPO in EXO 0748--676 with a frequency varying
between 0.58 and 2.44 Hz, and with rms amplitudes between 8.4\% and
12.1\%. The QPO was detected throughout the persistent emission, the
dips, and the type I X-ray bursts, with strengths that were consistent
with each other, except in observation 10, where the QPO was stronger
during the dips than in the persistent emission. Over the whole
frequency range $Q$ was consistent with 3.5, except in the X-ray bursts
where it was higher. The QPO was not detected in observation 1, when
the count rate was a factor 2 to 3 higher than otherwise.  The power
spectral features show no changes that are correlated with variations
of $\sim$20\% in the out-of-dip count rate, except for observations 8
to 10, in which the frequency is anti-correlated with the out-of-dip
count rate.  In addition to the QPO, a noise component at frequencies
below 1 Hz was present, with a typical strength of 6\%--10\% rms. Both
the QPO and the noise component have a flat photon energy spectrum.

Several of the properties of the QPO, notably its frequency, its
relatively unchanged persistence during bursts and dips, and its flat
energy spectrum, are remarkably similar to those of the $\sim$1 Hz QPO
that was recently found by JO99 in 4U 1323--62. Figure \ref{ratio}
shows that the ratio of frequency to FWHM of the QPO in 4U 1323--62 is
also consistent with 3.5, and Figure \ref{noise} shows that the noise
component in 4U 1323--62 falls on the strength versus slope relation
found for the noise component in EXO 0748--676. We conclude that both
QPOs are produced by the same mechanism. Since QPOs like these have
not been seen in other neutron star LMXBs the origin is most likely
related to the high inclination of the sources (JO99). It is unlikely
that the QPOs are produced by variations in mass accretion rate
($\dot{M}$) onto the neutron star: in that case they should also be
visible in low-inclination sources. Moreover, other QPOs that are
thought to be produced by $\dot{M}$ variations, show considerable
energy dependence (van der Klis \markcite{va95}1995). A medium
modulating the radiation from a central source (JO99) seems to be a
more promising explanation (see below).

We have found several new features of the QPO, that were not seen in
4U 1323--62. First,  the large flux range in the observations of EXO
0748--676 allow us to say that the QPO is not present at high
persistent count rates. In 4U 1323--62, whose flux varied by only
$\sim$15\%, the QPO was always present.  Second, the properties of the
QPO changed during three of the four bursts. In 4U 1323--62 the
properties of the QPO remained the same during all bursts. Third,
there is the difference in frequency range over which the QPOs are
observed: a factor $\sim$4 in EXO 0748--676 and $\sim$12\% in 4U
1323--62. This could be due to the difference in the time span over
which the sources were observed. 4U 1323--62 was observed for two
consecutive days, whereas the EXO 0748--676 observations span 
19 months.  The $\sim$12\% range in two days, is similar to the
$<$15\%  range found for individual observations
($\sim$1 day) of EXO 0748--676. Finally, the noise has a flat
photon spectrum, which suggests that the mechanism behind the noise is
similar to that for the QPO.

Much can be learned about the QPO from its behavior during the X-ray
bursts. The fact that the QPO is detected during all four bursts shows
that the presence of the QPO does not directly depend on the
instantaneous flux, and that the absence of the QPO in observation 1
is due to something else than just a higher count rate. Since the
fractional rms amplitudes during the bursts are similar to those
outside the bursts, variations in $\dot{M}$ onto to neutron
star can be ruled out as the origin of the QPO. A dramatic decrease in
rms amplitude is expected in that case. The high count rates during
the X-ray bursts enables us to detect the QPO in smaller time
intervals, typically tens of seconds, than otherwise
possible. In all bursts the $Q$ of the QPOs is higher than 3.5, and
during three of the bursts we see the frequency shift. This seems to
indicate that the QPO is intrinsically narrower than measured by us. The
observed width of the QPO outside the bursts is then due to changes of
the QPO frequency on time scales of a few hundred seconds, and/or due
to the presence of several QPOs with a higher $Q$ that occur around a
central frequency.

We now consider models that involve modulation of radiation from a
central source. Given the inclination of EXO 0748-676
(75$^\circ$--82$^\circ$), to be in the line of sight to the central
source, a medium modulating the radiation from the central source must
at least reach a height that is between 14\% and 27\% of its radial
distance to the central source. De Jong, van Paradijs, \& Augusteijn
(\markcite{devaau96}1996) found that accretion discs in LMXBs have
opening semi-angles of $\sim$12$^\circ$ as seen from the central
source. Hence only a small height above the accretion disc is
required, typically a few percent of the radial distance to the
neutron star, to be in the line of sight. This suggests that the
modulating medium is in or on the disc. Any model is constrained by
the fact that the rms amplitude stays constant when the source goes
into a dip; the medium causing the dips has to remove more or less the
same amount of modulated as unmodulated radiation from our line of
sight. Models can be constructed, with varying geometries that depend
on whether the central source is point-like or extended, and whether
the media causing the dips and QPO are absorbing or scattering
media. Only a few combinations can be ruled out with certainty,
e.g. most models involving a point-like source. If the dips are being
caused by a partial covering opaque medium, the central source has to
be extended, and the medium causing the QPO should have an azimuthal
extent comparable to that of the medium causing the dips, in order to
explain the constancy of the rms amplitudes in- and outside the dips.
The mechanism also has to produce a central frequency that does not
vary by more than 15\% over the duration of a single observation
(typically one day).

Both partial covering by an opaque medium (see JO99) and Thomson
scattering can explain the flat photon spectrum of the QPO. Assuming
the QPO frequency is the Kepler frequency of a structure orbiting
a 1.4 M$_\odot$ neutron star, one derives radial distances of
1.0--2.4$\times10^8$ cm. The viscous timescale (Frank, King, \& Raine
\markcite{frkira92}1992) at these radii, on which the orbiting
structure is expected to alter, is comparable to the timescales that
are seen for the frequency shifts during the burst. The larger
frequency changes between observations might be related to the
timescale on which the accretion disc as a whole changes its
structure (typically days to weeks). We remark that the
inferred radii are similar to those in the $\alpha$-disc model
(Frank et al.\markcite{frkira92}1992), at which the dominant opacity
source changes from Kramers' opacity to electron scattering. This
radius increases with $\dot{M}$, and implies an anti-correlation
between frequency and count rate, as seen in observation~8--10.

The fact that the QPO is not found in observation 1 suggests that the
higher count rate is accompanied by a change in the accretion disc
structure. Both are probably due to an increase in $\dot{M}$. Assuming
that an accretion disc thickens with $\dot{M}$, the modulated
flux might be obscured by a thicker disc, but the mechanism may also
have disappeared altogether.

Finally we remark that the 0.15 s jitter in the mid-eclipse timings
reported by Hertz et al. (\markcite{hewoco97}1997) might be due the
presence of the QPO reported here. A QPO with a frequency of 1 Hz and
an rms amplitude of 10\%, arbitrarily superimposed on eclipse
transitions lasting a few seconds, can be expected to cause a jitter in
the order of 0.1 s.

\acknowledgments{This work was supported in part by the Netherlands
Foundation for Research in Astronomy (ASTRON).  JVP acknowledges NASA
support.  This research has made use of data obtained through the
HEASARC Online Service, provided by the NASA/GSFC.}

\clearpage

\psfig{figure=two-qpo.ps,width=8cm}
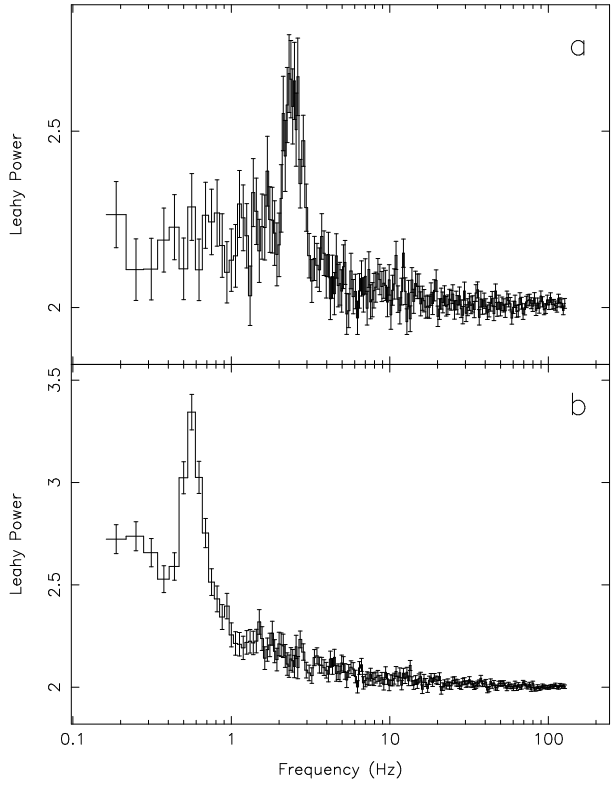
\figcaption[two-qpo.ps]{The power spectra of observations 2 (a) and 10
(b), showing the QPO at 2.44$\pm$0.03 Hz and at 0.58$\pm$0.01 Hz,
respectively.\label{qpo}}

\newpage
\psfig{figure=qpo-fwhm.ps,width=8cm}
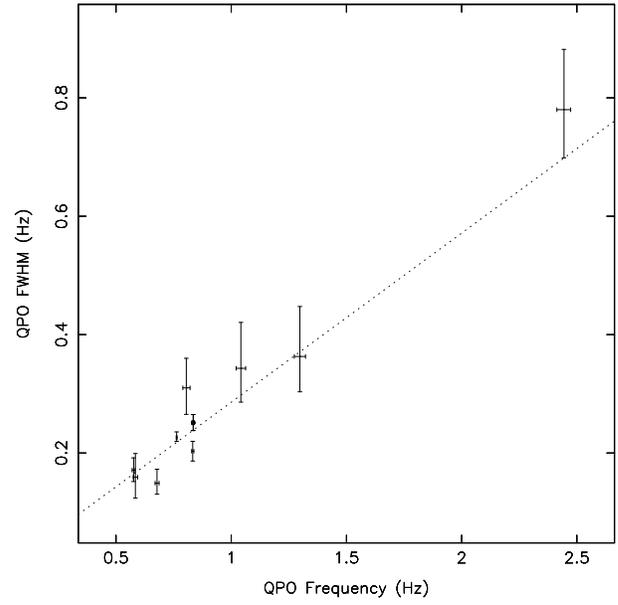
\figcaption[qpo-fwhm.ps]{The QPO FWHM as a function of the QPO
frequency. The dotted line is the best fit for a constant $Q$
($Q$=3.5). The black dot represents the value obtained for 4U 1323--62
(JO99).\label{ratio}}

\newpage
\psfig{figure=noise.ps,width=8cm}
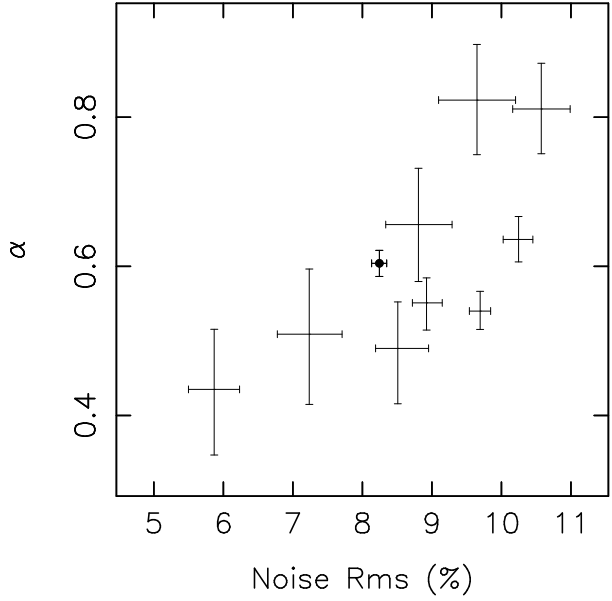
\figcaption[noise.ps]{The power law index, $\alpha$, of the noise
component as a function of the noise rms amplitude. The black dot
represents the value obtained for 4U 1323--62.\label{noise}}

\newpage
\psfig{figure=energy_depend_qpo.ps,width=8cm}
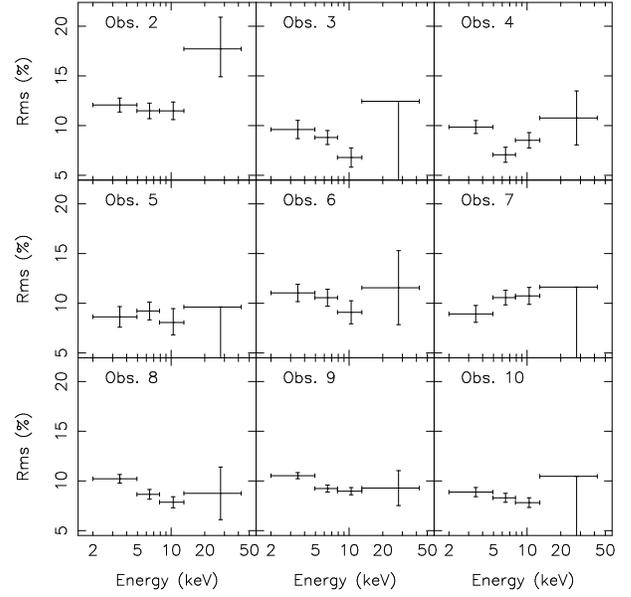
\figcaption[energy_depend_qpo.ps]{The energy dependence of the QPO
strength (dip and non-dip data combined). The points without positive rms errors are upper
limits.\label{energy}}

\clearpage

\begin{deluxetable}{ccccccccccc}
\tablecolumns{11}
\tiny
\tablewidth{0pt}
\tablecaption{Log of the {\it RXTE} observations of EXO 0748--676, and the
power spectral fit parameters for each of them.  The fit parameters are for all data (dip and non-dip) combined. For each burst the QPO parameters are given below the
observation in which they occurred; for burst 4 the fit parameters of both Lorentzians are given. \label{table1}}
\tablehead{
Obs.$^a$ & Begin           & End            & $T_{OBS}$$^b$ &        \multicolumn{3}{c}{QPO}	                                      &  & \multicolumn{2}{c}{Noise}	    &  Count rate$^c$          \\ \cline{5-7} \cline{9-10}
     &                 &                &           & rms                    & FWHM                    & Frequency              &  & rms$^d$ 	   & PLI	    &                      \\
     & (UTC)           & (UTC)          & (ks)      & (\%)                   & (Hz)                    & (Hz)                   &  & (\%)	   &	    &  (counts s$^{-1})$  
}
\startdata
1    &  12-03-96 10:41 & 12-03-96 12:54 & 6.0  (0.9)  &  $<2.1^e$	             &  ---          &  ---	        &  & 3.8$\pm$0.2	   & 0.98$\pm$0.12       & 177.2$\pm$0.3  \\
2    &  03-05-96 08:37 & 04-05-96 00:30 & 12.0 (2.9)  &  12.1$^{+0.6}_{-0.6}$  & 0.8$\pm$0.1             & 2.44$\pm$0.03          &  & 5.7$\pm$0.4	   & 0.43$\pm$0.09 & 72.9$\pm$0.3         \\
3    &  18-01-97 17:14 & 19-01-97 13:18 & 9.6 (3.7)   &  8.8$\pm$0.7           & 0.16$\pm$0.04           & 0.58$\pm$0.01          &  & 9.6$\pm$0.6	   & 0.82$\pm$0.07 & 62.7$\pm$0.2         \\
4    &  07-03-97 20:03 & 08-03-97 12:40 & 9.8 (5.6)   &  8.9$\pm$0.5           & 0.15$\pm$0.02           & 0.68$\pm$0.01          &  & 10.6$\pm$0.4	   & 0.81$\pm$0.06 & 59.6$\pm$0.2         \\
5    &  01-05-97 11:52 & 02-05-97 07:25 & 9.5 (3.3)   &  9.3$^{+0.8}_{-0.6}$   & 0.36$\pm$0.07           & 1.30$\pm$0.02          &  & 7.2$\pm$0.5	   & 0.51$\pm$0.09 & 64.9$\pm$0.3         \\
     &  Burst 1        &                &             &  9.7$\pm$0.9           & 0.18$\pm$0.5            & 1.41$\pm$0.02          &  &		   &	    &                      \\
6    &  26-06-97 06:01 & 26-06-97 22:03 & 9.7 (2.7)   &  9.8$^{+0.8}_{-0.6}$   & 0.34$\pm$0.07           & 1.04$\pm$0.02          &  & 8.8$\pm$0.5	   & 0.67$\pm$0.08 & 60.9$\pm$0.2         \\
7    &  07-10-97 00:51 & 07-10-97 20:32 & 8.6         &  9.9$\pm$0.6           & 0.31$\pm$0.05           & 0.81$\pm$0.02          &  & 8.5$^{+0.4}_{-0.3}$ & 0.49$\pm$0.07 & 68.3$\pm$0.3         \\
8    &  13-08-97 19:38 & 14-08-97 03:02 & 17.3 (1.7) &  9.1$\pm$0.3           & 0.20$\pm$0.02           & 0.833$\pm$0.005        &  & 8.9$\pm$0.2	   & 0.55$\pm$0.04 & 69.0$\pm$0.1         \\
     &  Burst 2        &                &             &  8.4$\pm$1.1           & 0.2$\pm$0.1             & 0.84$\pm$0.02          &  &		   &	    &                      \\
9    &  17-08-97 12:41 & 17-08-97 18:59 & 36.7 (6.9)  &  10.4$\pm$0.2          & 0.23$\pm$0.01           & 0.763$\pm$0.002        &  & 9.7$\pm$0.2	   & 0.54$\pm$0.03 & 72.5$\pm$0.1         \\
     &  Burst 3        &                &             &  7.1$\pm$0.9           & 0.11$\pm$0.03           & 0.75$\pm$0.02          &  &		   &	    &                      \\
     &  Burst 4        &                &             &  7.3$\pm$1.3           & 0.11$\pm$0.07           & 0.89$\pm$0.02          &  &		   &	    &                      \\
     &                 &                &             &  6.7$^{+2.8}_{-1.6}$   & 0.13$^{+0.17}_{-0.06}$  & 0.65$^{+0.03}_{-0.05}$ &  &		   &	    &                      \\
10   & 22-08-97 00:13  & 22-08-97 18:56 & 24.5 (10.1) &  8.4$\pm$0.4           & 0.17$\pm$0.02           & 0.58$\pm$0.01          &  & 10.2$\pm$0.2	   & 0.64$\pm$0.03 & 75.3$\pm$0.2         \\
\enddata
\tablenotetext{a} {Obs. IDs are 10068--03--01--00, 10108--01--(01--05), 20069--(02--06), and 20082--01--(01--03) for respectively obs. 1, 2, 3 to 7, and 8 to 10.}
\tablenotetext{b} {Time in dips (see section 2) is given between brackets. For observation 7 this could not be determined accurately.} 
\tablenotetext{c} {The average  2--60 keV count rate (3 detectors; dips, eclipses and bursts excluded), obtained from the Standard 2 data.}
\tablenotetext{d} {The noise rms is integrated from 0.1 to 1.0 Hz.}
\tablenotetext{e} {For the frequency range  1--10 Hz and for an assumed $Q$ of 3.5}
\end{deluxetable}

\end{document}